\documentclass{article}

\usepackage{arxiv}
\usepackage{subfigure}
\usepackage{adjustbox}
\usepackage{lineno,hyperref}
\modulolinenumbers[5]
\usepackage{algpseudocode}
\usepackage{graphicx}
\usepackage[most]{tcolorbox}

\usepackage{amsmath,amssymb,amsfonts}
\usepackage{amsthm}
\usepackage{booktabs}
\theoremstyle{definition}

\usepackage{url}
\usepackage{enumitem}
\setlist{leftmargin=5mm}
\usepackage{xcolor}
\usepackage[ruled]{algorithm2e}

\SetCommentSty{colortcp}
\usepackage{mathtools}
\DeclarePairedDelimiterX{\infdivx}[2]{(}{)}{#1\;\delimsize\|\;#2}
\newcommand{\infdiv}{KL\infdivx}

\usepackage{color}

\title{University of Washington at TREC 2020 Fairness Ranking Track}

\author{
    Yunhe Feng$^{1}$, Daniel Saelid$^{2}$, Ke Li$^{1}$, Ruoyuan Gao$^{3}$, Chirag Shah$^{1}$ \\
    $^{1}$Information School, University of Washington\\
    $^{2}$Paul G. Allen School of Computer Science \& Engineering, University of Washington\\
    $^{3}$Department of Computer Science, Rutgers University\\
    \texttt{\{yunhe,saeliddp,kel28,chirags\}@uw.edu} $\qquad$ \texttt{ruoyuan.gao@rutgers.edu}\\
}

\begin{document}

\maketitle


\begin{abstract}
InfoSeeking Lab's FATE (Fairness Accountability Transparency Ethics) group at University of Washington participated in 2020 TREC Fairness Ranking Track.
This report describes that track, assigned data and tasks, our group definitions, and our results.
Our approach to bringing fairness in retrieval and re-ranking tasks with Semantic Scholar data was to extract various dimensions of author identity.
These dimensions included gender and location.
We developed modules for these extractions in a way that allowed us to plug them in for either of the tasks as needed.
After trying different combinations of relative weights assigned to relevance, gender, and location information, we chose five runs for retrieval and five runs for re-ranking tasks.
The results showed that our runs performed below par for re-ranking task, but above average for retrieval. 
\end{abstract}


\section{Introduction}
\label{sec:introduction}

As one of the emerging topics in fairness-aware information systems, presenting relevant results to the users while ensuring fair exposure of the content suppliers have raised more and more attention.
Fairer information retrieval and search systems not only provide relevant search results with higher diversity and transparency, but also offer reasonable discoverability for underrepresented groups.
For example, a high-quality academic paper from small institutions and universities, which have very limited media outlets and resources, should also be treated equally to get its deserved exposures in search systems, especially at the early stage of publication when such papers are more likely to suffer from cold-start problems.

The TREC 2020 Fairness Track~\cite{trec-fair-ranking-2020,trec-fair-ranking-2019} initiated the task of fairness ranking within an academic search task context, where the goal was to provide fair exposure of different groups of authors while maintaining good relevance of the ranked papers regarding given queries. 
However, it is difficult to achieve such a goal due to the following challenges.

\begin{itemize}
    \item\textbf{Openness and complexity of defining the author group.}
    Defining the author group is not a trivial task.
    This requires an in-depth understanding of what should be considered as important group attributes that not only separate different authors but also aggregate similar authors.
    The challenges in this task include and are not limited to, how many groups should be identified, how to identify and extract the features from authors and their publications for the group classification task, and what algorithm to use for effective and efficient classification. 
    
    \item \textbf{Algorithm Robustness on different applications.}
    The definition of author groups may change from application to application.
    A good fairness ranking algorithm should be robust to a broad range of group definitions in various scenarios.
    In other words, fairness-aware ranking algorithms should demonstrate a high generalization capability when processing application-wise group definitions.
    
    \item \textbf{Dynamics of information retrieval given varying group definitions.}
    The task of retrieving relevant documents requires a good representation of each group. It is still an open question of how to effectively select a pool of relevant candidates that also covers a broad range of author groups with varying group definitions.
    
    \item \textbf{Trade-off between relevance and fairness.}
    The re-ranking algorithm based on a list of candidate items needs to optimize for both the relevance of the re-ranked results and the fairness of the exposed author groups, while carefully balancing between the two. 
\end{itemize}

At TREC 2020 fairness ranking track, we aimed to design and implement fairly ranking and retrieval algorithms to enhance the fairness for scholarly search.
On the subset of the Semantic Scholar (S2) Open Corpus\footnote{\url{http://api.semanticscholar.org/corpus/}}~\cite{ammar:18} provided by the Allen Institute for Artificial Intelligence, we defined multiple author groups, inferred demographic characteristics of authors, and developed fairness-aware algorithms to achieve a flexible trade-off between relevance and fairness by tuning principal component weights.
Specifically, we participated in two tasks: (1) retrieval task where the goal was to return a ranked list of papers to serve as the candidate papers for re-ranking; (2) re-ranking task where the goal was to rank the candidate papers based on the relevance to a given query, while accounting for the fair author group exposure.

\section{Data Description and Storage}
The Semantic Scholar (S2) Open Corpus used in TREC 2020 fairness ranking track consists of extracted fields of academic papers.
For most papers, the available fields include the S2 paper ID, title, abstract, authors, inbound and outbound citations.
The S2 corpus subset available to participants for training and testing consisted of 8,879 academic papers in total.
A given paper was associated with a unique S2 identifier, which we used as a key for querying purposes.
Beyond the identifier, a given paper could have non-empty data for any subset of the set of fields found at \url{http://api.semanticscholar.org/corpus/}~\cite{ammar:18}.
For re-ranking purposes, the following fields are likely most pertinent: title, abstract, author list, in-citations, out-citations, fields of study, journal name, journal volume, year of publication, and venue of publication.
Authors in the author list are identified by name and one or more unique author IDs.
The lists of in-citations and out-citations are simply collections of S2 identifiers referring to other papers in the sub-corpus.

\begin{figure}[ht]
  \centering
  \subfigure[Authors JSON data \label{fig:type-authors}]{\fbox{\includegraphics[width=0.33\linewidth]{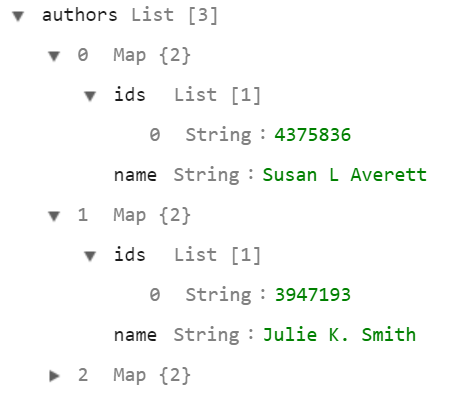}}}
  \hspace{1cm}
  \subfigure[In-Citations JSON data \label{fig:type-incitations}]{\fbox{\includegraphics[width=0.492\linewidth]{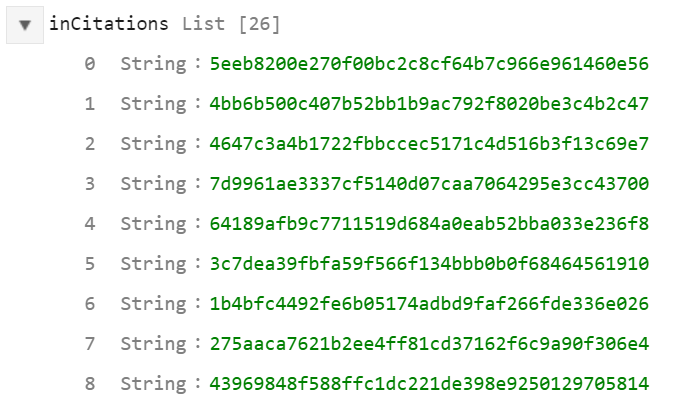}}}
  \caption{Snapshots of data stored on DynamoDB.\label{fig:DynamoDB}}
\end{figure}

We chose to upload the full sub-corpus to an AWS DynamoDB database for querying and processing.
This NoSQL approach was appropriate because we only needed to query the documents by S2 identifiers.
Figure~\ref{fig:DynamoDB} shows the snapshots of Authors JSON data, and In-Citations JSON data returned by Amazon DynamoDB during a query operation.

The TREC organizers also provided three CSV files.
One consisted of paper ids and a list of corresponding author positions with their corpus id.
Another one contained paper information such as paper id, title, year of publication, venue, number of citations, and number of key citations.
The other one mapped author IDs to a few features: author's name, number of citations, h-index (and a dependent feature, h-class), i10-Index, and number of papers published.

For the 31,975 authors included in the sub-corpus of papers, we constructed a similar DynamoDB database.
Using these two databases in tandem, we could extract the author group characteristics for a given paper.

We also received a training sample with the same JSON format as the evaluation data for training purposes.
Each JSON line contained the query id, query name, frequency, and a list of document ids.
For each document in the document list, we had its document id and a binary relevance score (0 or 1).
As described by TREC, the relevance score was extracted from real-time traffic.
That means if a person queries the corpus and clicks on one of the results, that selected document's relevance score will be set to 1.
And the other documents that are not being chosen will be considered irrelevant to the query. 
The evaluation data has the same format except that we do not have the relevance score for each document in the evaluation data.

\section{Methodology}
We first defined author groups based on general demographic characteristics including genders and countries (see below for details).
Then, we utilized Okapi BM25~\cite{bm25} to estimate the relevance of papers to given search queries.
Based on the group definition and BM25 relevance score, we proposed our fairness-aware re-ranking algorithm.

\subsection{Group Definition}
When defining author groups, we considered genders and countries of authors because the two demographic features are general enough for different applications.
Re-ranking algorithms based on such group definitions are more likely to demonstrate strong robustness in various scenarios.

\subsubsection{Gender Inference}
To predict the binary gender of a given author, we called the genderize.io API\footnote{\url{https://genderize.io/}}~\cite{genderize}, which is powered by a large dataset that maps first names to binary genders.
Given a name, genderize.io will return `male' if there are more instances of the name associated with men, and it will return `female' otherwise.
If the dataset contains no instances of the given name, no gender prediction will be returned.
For the authors in our sub-corpus, the returned gender predictions are shown in Table~\ref{tab:inferred_gender}.

\begin{table}[]
    \begin{center}
        \caption{The distribution of inferred genders by genderize.io} 
        \label{tab:inferred_gender} 
        \begin{tabular}{l|l|l}
        \toprule
        Gender       & Count & Percentage \\ \midrule
        Male         & 18810 & 58.8\% \\ 
        Female       & 6235  & 19.5\% \\ 
        Unidentified & 6930  & 21.7\% \\ \midrule
        Total        & 31975 & 100\% \\ \bottomrule
        \end{tabular}
    \end{center}
\end{table}

\subsubsection{Country Inference}
In contrast with gender prediction, we could not rely on a single API call for location prediction.
To begin the process, we searched for the author by name in Google Scholar using the Scholarly API\footnote{\url{https://pypi.org/project/scholarly/}}~\cite{scholarly-lib}.
Since there are often many authors with a given full name on Google Scholar, we picked a single author by comparing our author citation data with Google Scholar's data.
After choosing the closest match, we then retrieved email extension and `affiliation' data from Google Scholar.
If we successfully retrieved this author data, we followed the below procedure, moving to each consecutive step if the prior was unsuccessful.
As listed as the last step, if no author data was retrieved from Google Scholar, we tried finding the author's homepage and parsing its URL for country code.

\begin{enumerate}
  \item Parse the email extension for a country code (e.g. .uk $\xrightarrow{}$  the United Kingdom).
  \item Parse the affiliation for a university name, then return the country in which that university is located.\footnote{\url{https://www.4icu.org/reviews/index0001.htm}}
  \item Parse the affiliation for a city name, then return that city's country.\footnote{\url{https://en.wikipedia.org/wiki/List_of_towns_and_cities_with_100,000_or_more_inhabitants/cityname:_A}} 
  \item Search author name, author affiliation on Google, scrape the first URL, then parse for country code.
  \item Call Google Places API with author affiliation, then return associated country.
  \item Search author name + `homepage' on Google, scrape the first URL, then parse for country code.
\end{enumerate}

Once all authors had been run through this process, we mapped each author's affiliated country to `advanced economy' or `developing economy' based on the IMF's October 2019 World Economic Outlook report~\cite{world-economic-outlook}.
The results are shown in Table~\ref{tab:inferred_location}. Here, `unidentified' means that no country was predicted for that author.

\begin{table}[]
    \begin{center}
        \caption{The economy distribution of inferred locations} 
        \label{tab:inferred_location} 
        \begin{tabular}{l|l|l}
        \toprule
        Locations       & Count & Percentage \\ \midrule
        Advanced         & 15106 & 47.2\% \\ 
        Developing       & 3926  & 12.3\% \\ 
        Unidentified & 12933  & 40.5\% \\ \midrule
        Total        & 31975 & 100\% \\ \bottomrule
        \end{tabular}
    \end{center}
\end{table}

\subsection{Pure Relevance with BM25}
As one of the most popular ranking algorithms, Okapi BM25 is adopted by many search engines to calculate and predict the relevance of a document based on a given query.
For both TREC tasks, retrieval and ranking, we used Gensim's BM25 module to predict the relevancy for each document in the provided document list. 

Before running BM25, we preprocessed the sub-corpus to extract useful information.
Since the complete paper for each document in the sub-corpus was not provided by TREC organizers, we instead chose the paper's abstract and title to represent the corresponding document.
The papers were written in 28 different languages (detected by the langdetect API\footnote{\url{https://pypi.org/project/langdetect/}}~\cite{langdetect,nakatani2010langdetect}) including English, Arabian, German, Chinese, etc., while all the queries were in English only.
However, BM25 functions are incompatible with certain languages that cannot be tokenized by whitespace.
Therefore, we decided to translate all needed documents into English first and stored the tokenized text in the database for further usage.

Then we started the BM25 process, which was similar for both retrieval and ranking tasks.
We first translated and tokenized the queries since some of them contained Unicode.
After that, for each query, we calculated the BM25 score as the base relevance score for each document in the baseline document list provided by TREC organizers, and then arranged the documents based on their scores in descending order.
This sorted list was used as the pure ranking list for the given query.

\subsection{Fairness-aware Re-ranking Algorithm}
We proposed a fairness-aware re-ranking algorithm incorporating both relevance and diversity of documents.
The main idea was to estimate the cost of adding a document to the rank list $\mathbf{R}$ from the perspective of relevance and fairness.
For a document of $d$, we used $F(d, \mathcal{D}, q)$, the reversed normalized BM25 score of $d$ in a corpus $\mathcal{D}$ given a query $q$, to represent its relevance cost, where $0$ corresponds to most relevant, and $1$ corresponds to least relevant.

For a given query $q$, we first retrieved the top relevant documents to build a candidate corpus $\mathcal{D'}$.
To ensure ranking fairness, it is intuitive to make the probability of defined groups over the rank list $R$ and the candidate corpus $\mathcal{D'}$ very similar.
Specifically, let $p(v, \mathcal{D})$ be the probability distribution of a discrete group variable $v$ over the the document corpus $\mathcal{D}$.
Based on our group definitions, $v$ could be either the group of \texttt{gender} $g$ or \texttt{country} $c$, i.e., $v\in{\{g,c\}}$.
Note that this is flexible to be extended to other group definitions.
Then we use the Kullback-Leibler (KL) divergence of the group distribution probability between the updated current rank list $\mathbf{R}$ and the whole candidate corpus $\mathcal{D'}$ to measure their similarities.
We also assigned weights $\mathbf{w}$ for relevance cost and fairness cost for each defined group.
The cost function is expressed as:

\begin{equation}
    C(d, \mathbf{w}, \mathbf{R}, \mathcal{D'}, q) = w_r * F(d, \mathcal{D'}, q) + \sum_{v\in{\{g,c\}}} w_v * \infdiv{p(v, \mathbf{R} + \{d\})}{p(v, \mathcal{D'})}
\label{equ:rank_cost}
\end{equation} 

where $\mathbf{w} = \{w_r, w_g, w_c\}$ and $w_r+w_g+w_c=1$;
$F(d, \mathcal{D'}, q)$ is the reversed normalized BM25 score of a document $d$ such that 0 corresponds to most relevant, and 1 corresponds to least relevant;
and $\infdiv{p(v, \mathbf{R} + \{d\})}{p(v, \mathcal{D'})}$ is the Kullback-Leibler divergence regarding group $v$ between the updated $\mathbf{R}$ by appending document $d$ and the overall candidate corpus $\mathcal{D'}$.
Then, we built our re-ranked list by repeatedly appending the document with the minimal cost $C(d, \mathbf{w}, \mathbf{R}, \mathcal{D'}, q)$.
The proposed fairness-aware re-ranking algorithm as illustrated in Algorithm~\ref{alg:ranking}.

Since many documents were missing group definitions for at least one author, we adopted a systematic way to address this missing data.
For every author missing a group definition, we assigned a group value based on the overall group distribution in the corpus.
For instance, if 75\% of the authors in the corpus were identified as male, we choose `male' for an unidentified author with a probability of 75\%.

\begin{algorithm}[h]
\textbf{Input}: $\mathcal{D}$: document corpus; $q$: query of interest; $l$: length of expected ranked list ; $\mathbf{w}$: component weight vector 

\textbf{Output}: $\mathbf{R}$: re-ranked list of relevant documents 

$\mathbf{R} \leftarrow \O$ \tcp*[l]{initialize the ranked list as empty} 
$\mathcal{D'}, \mathcal{D''} \leftarrow$ Retrieve relevant document candidates from $\mathcal{D}$ for query $q$ \tcp*[l]{document candidate corpus for $q$} 
\For{$i = 1 \to l$}{
    $c_{min} \leftarrow A\:Large\:Integer$\tcp*[l]{initialize the minimal cost}
    $d_{min} \leftarrow None$ \tcp*[l]{initialize the document with the minimal cost}
    \For{$d \in \mathcal{D''}$}{
        Calculate the cost $C(d, \mathbf{w}, \mathbf{R}, \mathcal{D'}, q)$ according to Equation~\ref{equ:rank_cost} \tcp*[l]{calculate the cost of adding $d$ into $\mathbf{R}$}
        \If{$C(d, \mathbf{w}, \mathbf{R}, \mathcal{D'}, q) < c_{min}$}{ 
            $d_{min} \leftarrow d$ \tcp*[l]{update the document with the minimal cost}
            $c_{min} \leftarrow C(d, \mathbf{w}, \mathbf{R}, \mathcal{D'}, q)$ \tcp*[l]{update the minimal cost}
        }
    }
    append $d_{min}$ to $\mathbf{R}$ \tcp*[l]{add the document with the minimal cost into the re-ranked list $\mathbf{R}$}
    $\mathcal{D''} \leftarrow \mathcal{D''} - \{d_{min}\}$ \tcp*[l]{remove the added document $d_{min}$ from $\mathcal{D''}$}
}
\Return $\mathbf{R}$
\caption{Fairness-aware Re-ranking Algorithm}
\label{alg:ranking}
\end{algorithm}

\section{Results and Discussion}

\begin{figure}[ht]
  \centering
  \subfigure[Country \label{fig:performance_country}]{\includegraphics[width=0.48\linewidth]{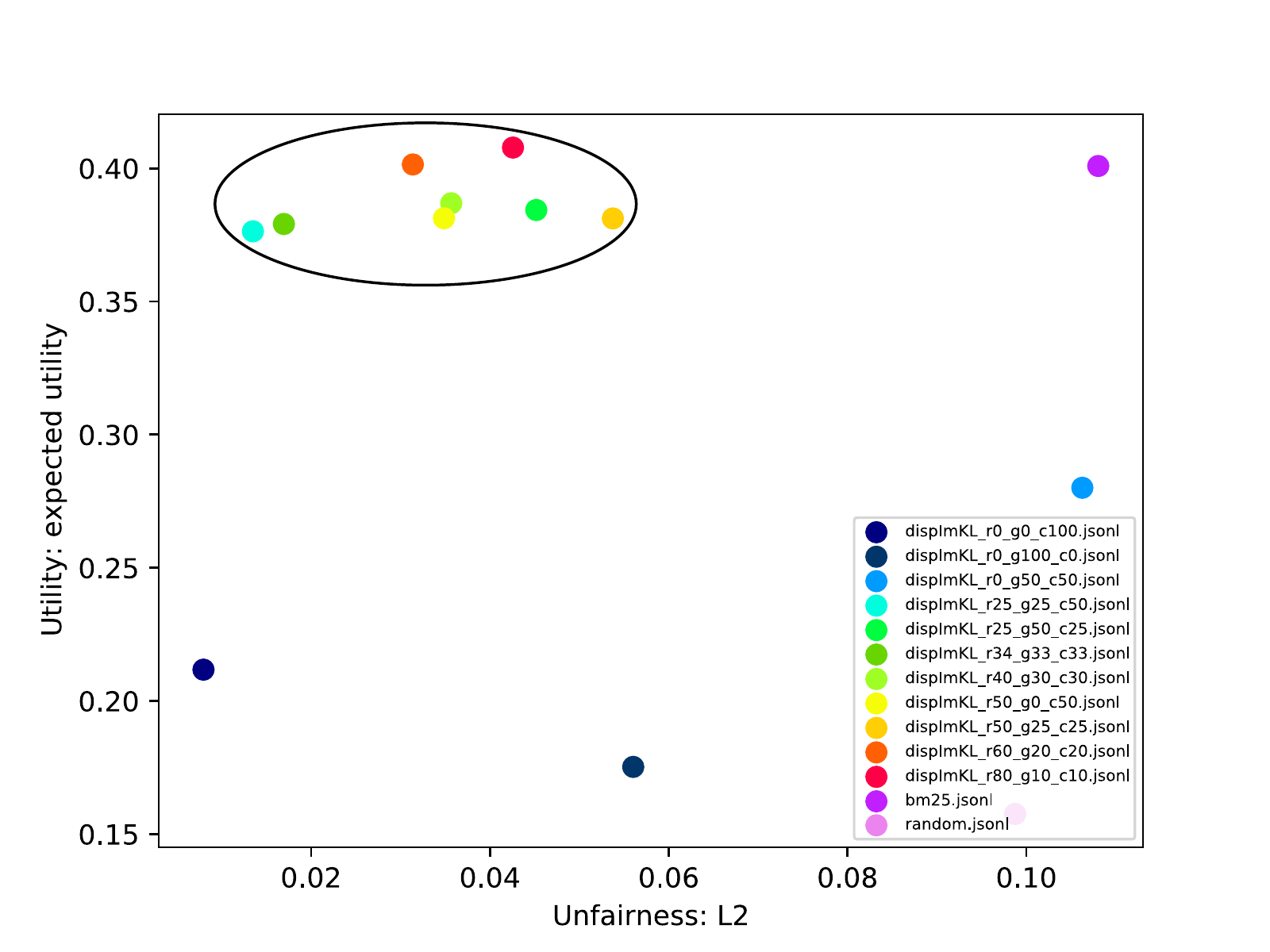}}
  \subfigure[Gender \label{fig:performance_gender}]{\includegraphics[width=0.48\linewidth]{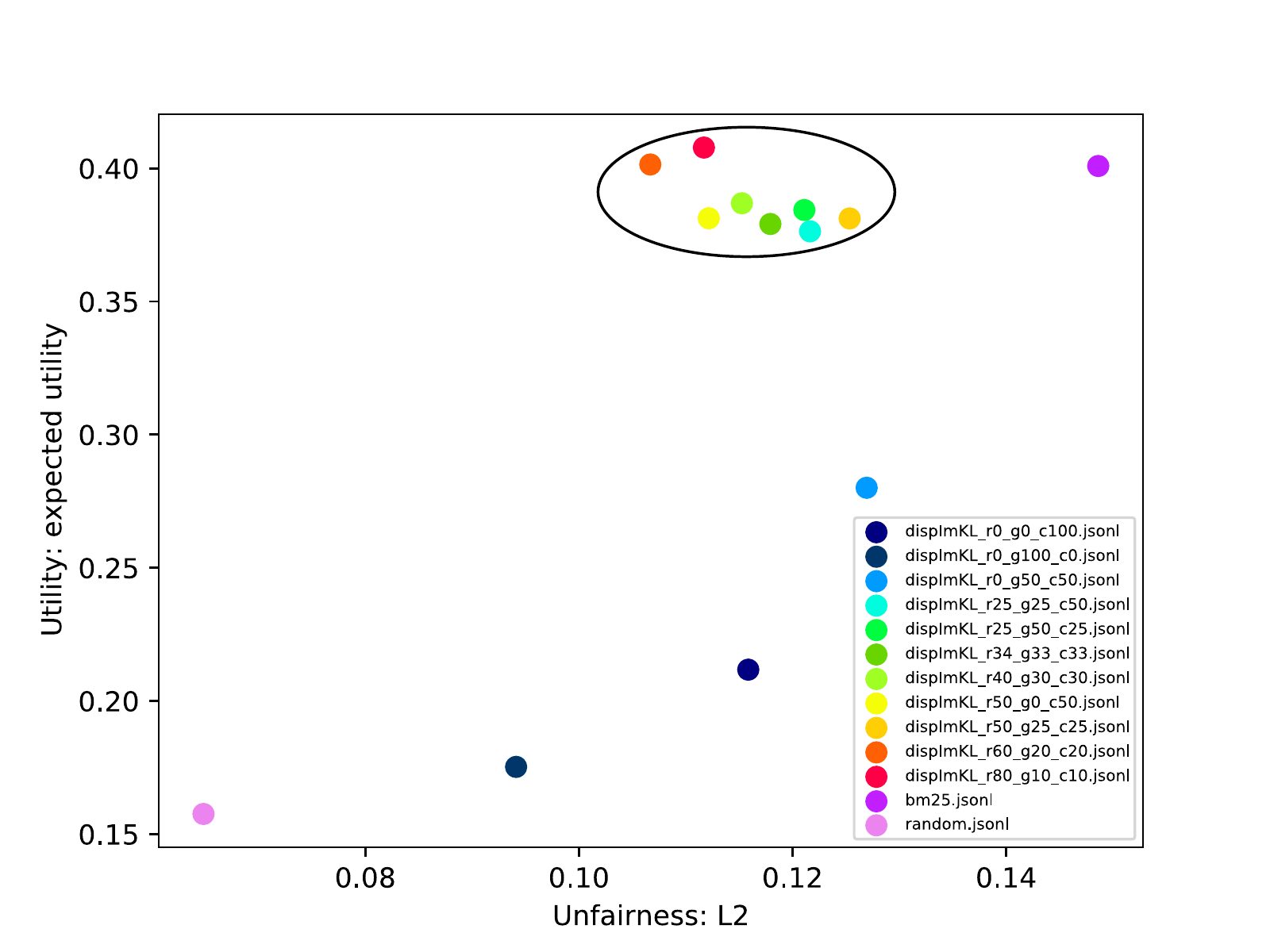}}
  \caption{Utility versus unfairness. \label{fig:performance}}
\end{figure}

We evaluated the utility and unfairness with different combinations of $w_r, w_g, w_c$ in Equation~\ref{equ:rank_cost} from the perspective of relevance, the group of gender, and the group of country, as shown in Figure~\ref{fig:performance}.
In both gender and country groups, BM25 demonstrates a relatively high utility score but a low fairness score, implying that BM25 fails to take fairness into account when calculating the ranking.
Another interesting finding is that the random ranking achieves lower fairness than most of our proposed methods on the country group but the highest fairness on the gender group.
So, the fairness performance of random ranking methods is sensitive to the definition of groups.
In other words, the definition of groups is not a trivial task as we claimed in Section~\ref{sec:introduction}.
As we expected, our methods' utility drops greatly when BM25 scores are excluded ($w_r = 0$).
When $w_r$ is assigned a positive value, the performance of our methods with different combinations of $w_r, w_g, w_c$ are comparable on both country and gender groups (see the cluster on left top in Figure~\ref{fig:performance_country}, and the cluster on the middle top in Figure~\ref{fig:performance_gender}).

\section{Future Work}
\begin{figure}[ht]
  \centering
  {\fbox{\includegraphics[width=0.8\linewidth]{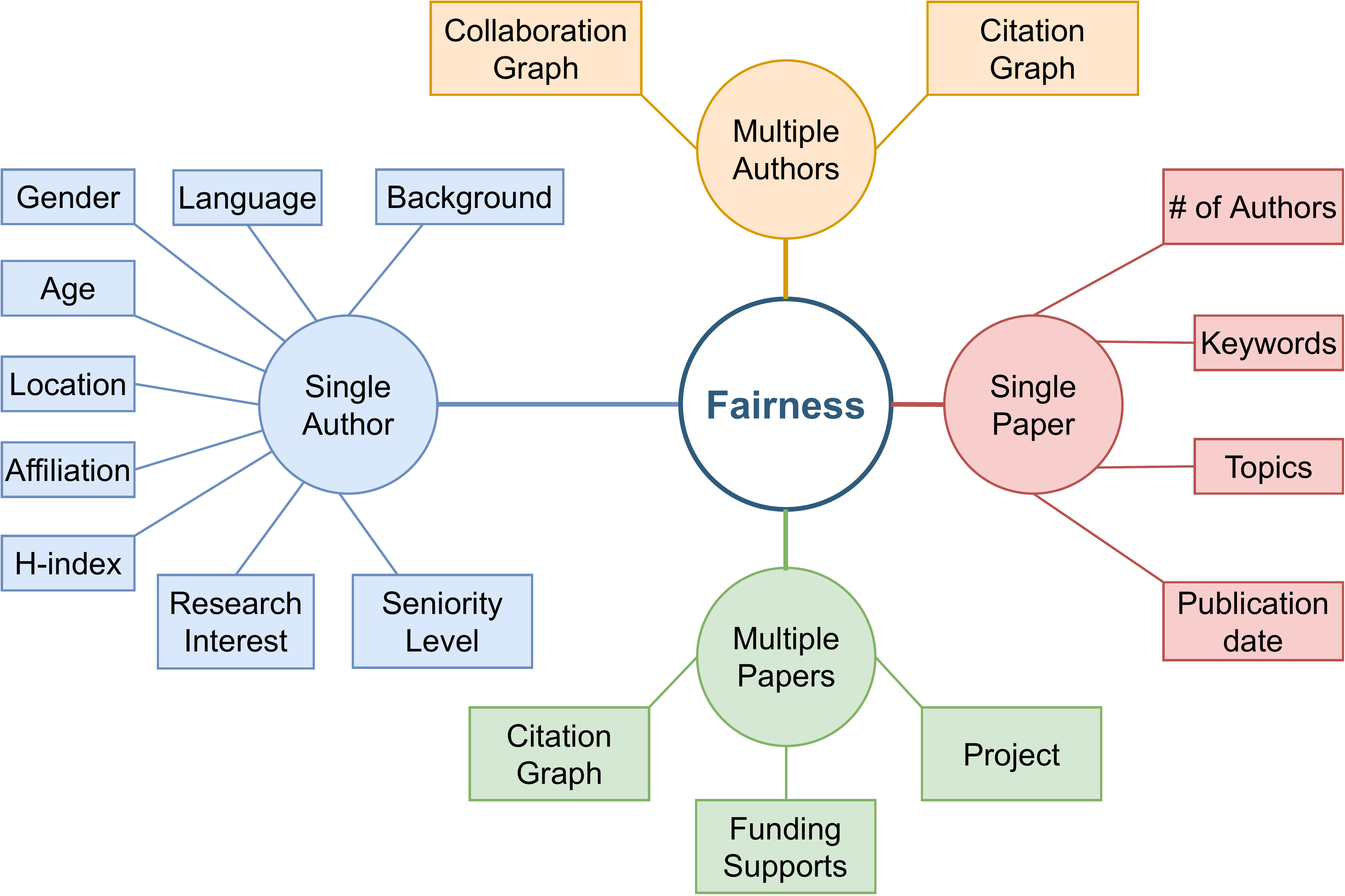}}}
  \caption{Fairness features to be explored
  \label{fig:fairness_features}}
\end{figure}

In the future, we will further explore more features of both authors and papers (see Figure~\ref{fig:fairness_features}) to improve the fairness in retrieval and re-ranking tasks with scholar data.
For a single author, we can consider both traditional demographic features including genders, ages, locations and languages, and the research related features such as affiliations, H-index, research interests, seniority levels (e.g., professors, postdocs, and PhD students), and backgrounds (e.g., academia, industry, and government).
When taking multiple authors into account, the citation and co-authorship historical data can be used to build up networking graphs, from which many measurements, such as centrality, will be calculated.
Adding the papers from authors who are far away from the influencers on the graphs may benefit diversity and fairness.

For the single paper, metadata such as keywords can be used to label the document. Besides, we can extract the topics from the paper abstract using topic models if possible. Other features, e.g., the number of coauthors, publication date, and total citations, can also be included.
It is common that multiple papers are supported by the same funding sources or derived from the same projects. For example, given the keyword of ``proactive searching'', most of the paper in search results probably come from the InfoSeeking Lab.
But it hurts diversity as the voices from small labs/universities and those who are not among the first to explore this research direction have less opportunity to be heard.
In addition, the citation graph enables the measurement of the closeness of multiple papers.

\section{Conclusion}
At TREC 2020 Fairness Ranking Track, InfoSeeking Lab's FATE (Fairness Accountability Transparency Ethics) group at University of Washington proposed a fairness-aware retrieval and re-ranking algorithm incorporating both relevance and fairness for Semantic Scholar data.
Evaluation results with different weights of relevance, gender, and location information demonstrated that our algorithm was flexible and explainable.
In this report, we also highlighted various challenges and possible directions for future work.

\section*{Acknowledgements}
A part of this work is supported by the US National Science Foundation (NSF) award number IIS-1910154.

\bibliographystyle{unsrt}
\bibliography{reference}

\end{document}